\documentclass{iopjournal}
\usepackage{lmodern}
\usepackage{amsmath}
\usepackage{amssymb}
\usepackage{graphicx}
\usepackage{cite}

\begin{document}

\articletype{Paper}%	 e.g. Paper, Letter, Topical Review...

\title{Enhanced charging power in nonreciprocal quantum battery by reservoir engineering}

\author{Qi-Yin Lin$^1$, Guang-Zheng Ye$^{1}$\orcid{{0009-0000-0521-0945}}, Can Li$^{1}$, Wan-Jun Su$^{1,*}$\orcid{0000-0003-1628-2279}, and Huai-Zhi Wu$^{1,*}$\orcid{0000-0002-2298-4819} }

\affil{$^1$Fujian Key Laboratory of Quantum Information and Quantum Optics, Department
of Physics, Fuzhou University, Fuzhou 350116, People's Republic of China}

\email{wanjunsu@fzu.edu.cn} 
\email{huaizhi.wu@fzu.edu.cn}


\begin{abstract}

We propose a scheme to achieve a nonreciprocal quantum battery (QB) in the non-Hermitian (NH) system, which can overcome the intrinsic dissipation and reverse flow constraints. The design is based on a charger and a battery, which are coherently coupled and jointly interact with a bad cavity. By introducing the auxiliary bad cavity and exploiting the nonreciprocal condition, this model can harness the environmental dissipation to suppress the reverse energy transfer. Under resonant conditions, we have achieved a four ratio of the battery energy to the charger energy; in contrast, this ratio is significantly reduced under large detuning. Through damping optimization, high efficiency of the short-time charging power is attained. In comparison to the fully nonreciprocal scheme, the QB operating at the exceptional point (EP) exhibits greater resilience to parameter fluctuations. These findings highlight the potential of NH quantum engineering for advancing QB technology, particularly in regimes involving directional energy transfer, controlled dissipation, and entropy management in open quantum systems.
\end{abstract}

\section{Introduction}

Recently, quantum batteries (QBs), first conceptualized by Alicki and Fannes in 2013 \cite{10.1103/PhysRevE.87.042123_2013} as a green solution employing quantum states for energy storage, have emerged as a pivotal direction in advanced battery technology\cite{10.1103/PhysRevA.109.062432_2024, 10.1103/d9k175d4_2025, 10.1103/PhysRevLett.134.180401_2025}. Entanglement and quantum coherence in QBs constitute essential resources during charging, offering performance advantages over classical counterparts\cite{10.1103/PhysRevLett.129.130602_2022}. Recent studies further indicate that collective solid-state charging can substantially enhance energy storage capacity \cite{10.1103/PhysRevLett.120.117702_2018}. However, the practical implementation of QBs still faces several key challenges, including decoherence caused by environmental disturbances, such as thermal fluctuations and electromagnetic noise, which degrades macroscopic quantum states and disrupts energy transfer \cite{10.1103/PhysRevLett.132.090401_2024, 10.1103/PhysRevA.102.060201_2020, 10.1088/1367-2630/ad3843,10.1209/0295-5075/ad2e79}. In this context, dissipative engineering has been proposed as a powerful approach to harness environmental dissipation as a beneficial resource rather than a drawback \cite{10.1088/0256307X/39/2/020502_2022}. For instance, decoherence suppression via periodic driving fields has extended coupling ranges beyond direct interactions, enabling waveguide-mediated long-distance charging\cite{10.1103/PhysRevLett.127.083602_2021}. Moreover, dissipative charging protocols based on the Dicke model have demonstrated charging power that scales as  $N^{2}$ \cite{10.1103/PhysRevLett.134.130401_2025}. These advances underscore the potential of engineered dissipation for overcoming existing limitations and advancing QB technologies toward real-world applications.

Non-Hermitian (NH) systems offer distinctive advantages for information transmission, largely due to the singular physical effects arising from their open-system characteristics governed by balanced gain and loss mechanisms \cite{10.1103/PhysRevApplied.21.L061002_2024, 10.1103/PhysRevA.110.062409_2024, 10.1103/PhysRevLett.122.253602_2019, arXiv:2411.19905_2024,10.1126/science.abl6571_2022}. The engineered violation of Lorentz reciprocity in such systems enables robust unidirectional signal propagation, which effectively suppresses back-reflection and associated interference \cite{10.1088/0256307X/42/4/047303_2025,10.7566/JPSJ.92.104705_2023,10.1038/ncomms3533_2013}. Within this framework, Ahmadi and Mazurek et al. recently introduced a design for a nonreciprocal QB \cite{10.1103/PhysRevLett.132.210402_2024}. Their scheme employs the interplay between coherent coupling and shared-reservoir interference effects to mitigate energy backflow and oscillatory losses observed in conventional battery models. Notably, this design remains efficient even under strong environmental dissipation, highlighting the potential of tailored non-Hermiticity in practical quantum energy architectures.

Inspired by the previous work, we propose an NH QB consisting of three bosonic modes. One strongly dissipative mode acts as an auxiliary cavity, while the other two, coupled to this cavity, serve as the charger and the battery. Through the projection operator method, we obtain an effective non-Hermitian Hamiltonian for the charger-battery subsystem. The nonreciprocity is achieved by balancing coherent and dissipative couplings, thereby transforming loss into directional energy transfer and significantly improving storage efficiency. Compared to two-level systems, bosonic QB offers higher capacity even under decoherence. Furthermore, we examine the QB working at an EP within the same framework, which is equivalent to a partially nonreciprocal configuration. A systematic comparison is then made with the fully nonreciprocal scenario. Experimentally, shared dissipation channels have been demonstrated in optomechanical and circuit-QED platforms \cite{10.1088/0256-307X/43/1/010302, 10.7498/aps.66.160302_2017, 10.1088/1402-4896/ad1a01, 10.1103/PhysRevA.109.012202_2024, 10.1088/1674-1056/adec5e, 10.1038/s4156702302128x_2023, 10.1038/s4158602407174w_2024}, supporting the feasibility of our scheme. This work suggests potential applications in quantum power systems, renewable energy storage, and compact energy devices.

\section{Theoretical Model}

\begin{figure}
\centering

\includegraphics[clip,width=0.8\columnwidth]{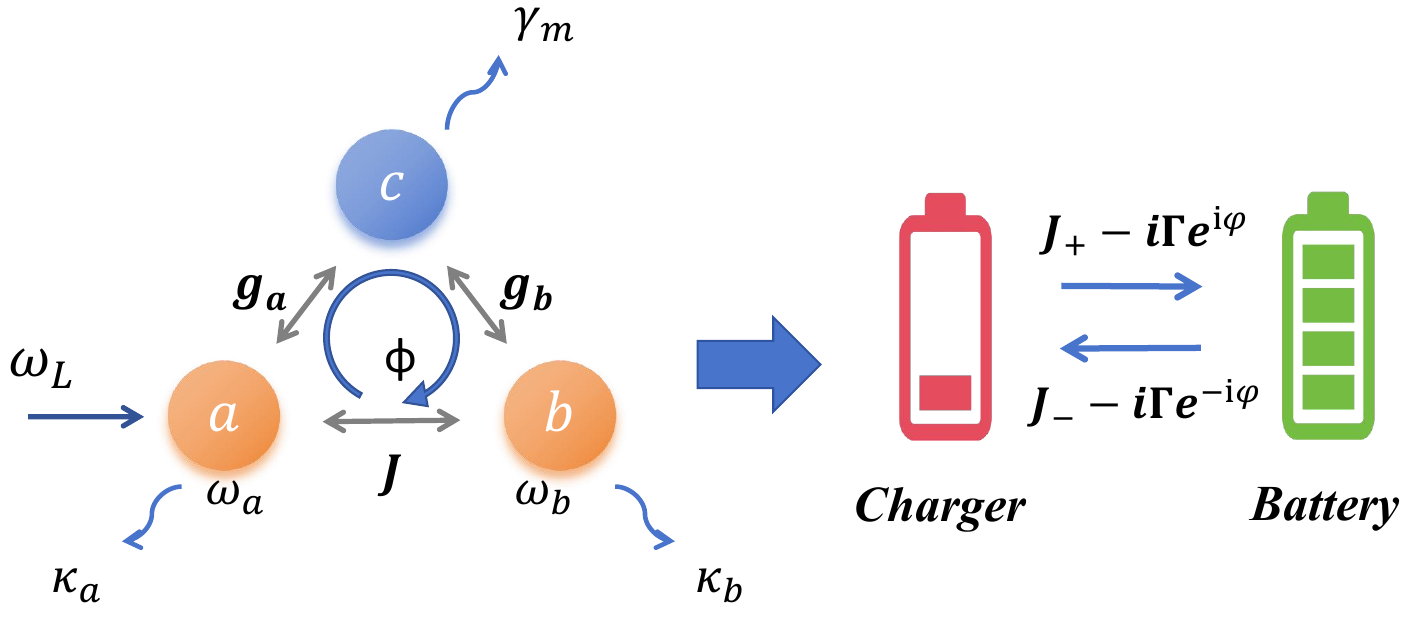}

\caption{\label{fig1}A schematic diagram of a non-Hermitian quantum battery system. The quantum charger $a$ (frequency $\omega_{a}$) interacts with the quantum battery $b$ (frequency $\omega_{b}$) with coupling strength $J$, and they collectively interact with mode $c$ with coupling strengths $g_{i}, i=a,b$, respectively. $\phi$ is the global phase between the three modes. The charger is driven by a field with amplitude $\varepsilon$ and frequency $\omega_{L}$. Here, $\kappa_{i}$ and $\gamma_{m}$ describe the damping rates of the respective modes. By projecting out mode $c$, the system will be equivalent to a direct interaction between mode $a$ and mode $b$, with a forward (backward) coupling rate $J_{+}-i\Gamma e^{i\phi}$  ($J_{-}-i\Gamma e^{-i\phi}$).}
\end{figure}

We consider the scenario where a strongly dissipative cavity mediates the charging process of a QB, as shown in Fig. \ref{fig1}. The system consists of three core components: a quantum charger and a quantum battery denoted by cavity $a$ and $b$, with the corresponding resonance frequency $\omega_{i}$ and damping rate $\kappa_{i}$ $(i=a,b)$, and a shared cavity (mode $c$) with resonance frequency $\omega_{c}$ and damping rate $\gamma_{m}$. For simplicity, we assume that both the charger and the battery have the same resonance frequency $\omega=\omega_{a}=\omega_{b}$. The coherent interaction between the charger and the battery is established with the coupling rate $J$. Simultaneously, both the charger and the battery are coupled to mode $c$ with coupling strengths $g_{i}$. The phase angle $\phi$ is the global phase between the three modes. The charger is pumped by an external coherent drive with amplitude $\varepsilon$ and frequency $\omega_{L}$. The system Hamiltonian can be expressed as (set $\hbar=1$):
\begin{eqnarray}
H & = & \Delta_{a}a^{\dagger}a+\Delta_{b}b^{\dagger}b+(\Delta_{c}-i\frac{\gamma_{m}}{2})c^{\dagger}c\nonumber \\
 &  & +g_{a}e^{i\phi}ac^{\dagger}+g_{a}e^{-i\phi}a^{\dagger}c+g_{b}(bc^{\dagger}+b^{\dagger}c)\nonumber \\
 &  & +J(ab^{\dagger}+a^{\dagger}b)+\varepsilon(a+a^{\dagger}),\label{eq:Hamiltonian}
\end{eqnarray}
where $a$ ($b$) and $c$ are the annihilation operators for the
charger (battery) and the bad cavity mode, respectively. $\Delta_{j}=\omega_{j}-\omega_{L}$ $(j=a,b,c)$ is the detuning between the corresponding mode and the driving field. 

When the dissipation rate of mode $c$ meets $\gamma_{m}\gg\{\kappa_{a},\kappa_{b}\}$, we can take mode $c$ as a mediator reservoir. This mediate coupling can be achieved by considering the shared slot as a damped cavity, waveguide, or transmission cavity \cite{10.1038/s4146701701304x_2017,10.1103/PhysRevLett.123.127202_2019, J.Appl.Phys.135_2024}. Under adiabatic conditions $|\Delta_{c}+i\gamma_{m}|\gg\{ g_{a},g_{b}\}$, we adiabatically eliminate the mode $c$, establishing an effective coupling between the charger and battery \cite{10.1103/PhysRevX.5.021025_2015}, see the right panel in Fig. \ref{fig1}. Using the projection operator method, $H_{\textrm{eff}}=PHP+\frac{PHQQHP}{E-QHQ}$, where the projection operators $P=\sum_{n}(|a_{n}\rangle\langle a_{n}|+|b_{n}\rangle\langle b_{n}|)$ and $Q=\sum_{n}|c_{n}\rangle\langle c_{n}|$ \cite{10.48550/arXiv.2201.09478_2022}. Assuming $E\ll |\Delta_{c}-i\gamma_{m}/2|$, the corresponding effective Hamiltonian projected within the subspace $\{|a_{n}\rangle,|b_{n}\rangle\}$ is rewritten as (\ref{Appendix:A}): 
\begin{eqnarray}
H_{\textrm{eff}} & = & H_{\textrm{C}}+H_{\textrm{D}},\nonumber \\
H_{\textrm{C}} & = & (\Delta_{a}^{\prime}-i\Gamma_{a})a^{\dagger}a+(\Delta_{b}^{\prime}-i\Gamma_{b})b^{\dagger}b\nonumber \\
 &  & +J_{-}a^{\dagger}b+J_{+}ab^{\dagger}+\varepsilon(a+a^{\dagger}),\nonumber \\
H_{\textrm{D}} & = & -i\Gamma e^{-i\phi}a^{\dagger}b-i\Gamma e^{i\phi}ab^{\dagger},\label{eq:Effective Hamiltonian}
\end{eqnarray}
where $\frac{\gamma_{m}}{2}=\gamma$,  $\Delta_{a}^{\prime}=\Delta_{a}-\frac{g_{a}^{2}\Delta_{c}}{\Delta_{c}^{2}+\gamma^{2}}$, 
$\Delta_{b}^{\prime}=\Delta_{b}-\frac{g_{b}^{2}\Delta_{c}}{\Delta_{c}^{2}+\gamma^{2}}$, 
$\Gamma_{a}=\frac{g_{a}^{2}\gamma}{\Delta_{c}^{2}+\gamma^{2}}$,  $\Gamma_{b}=\frac{g_{b}^{2}\gamma}{\Delta_{c}^{2}+\gamma^{2}}$, 
$G=\frac{g_{a}g_{b}\Delta_{c}}{\Delta_{c}^{2}+\gamma^{2}}$,  $\Gamma=\frac{\gamma g_{a}g_{b}}{\Delta_{c}^{2}+\gamma^{2}}$, 
and $J_{\pm}=J-Ge^{\pm i\phi}$. The entire system can be regarded as the interaction between modes $a$ and $b$, encompassing both coherent and dissipative coupling interactions.

For a Markovian reservoir, the reduced density matrix $\widehat{\rho}$ of the two cavity modes can be described using the Lindblad master equation with corresponding dissipators:
\begin{equation}
\dot{\rho}=-i\left[H_{\textrm{eff}},\rho\right]+\sum_{i=a,b}\kappa_{i}\mathcal{L}[i]\rho.\label{eq:The general master equation}
\end{equation}
The first term of the master equation describes the effective coupling between the charger and battery, and the second term $\mathcal{L}[o]$ represents the dissipative superoperator for modes $a$ and $b$ coupled to the environment, which is defined as
\begin{equation}
\mathcal{L}[o]\rho=o\rho o^{\dagger}-\frac{1}{2}\left\{ o^{\dagger}o,\rho\right\} .\label{eq:The standard dissipative superoperator}
\end{equation}
For $g_a=g_b$ and then $\Gamma_{a}=\Gamma_{b}=\Gamma$, the time evolution of the mean values of the cavity operators is governed by the following equations
\begin{eqnarray}
\dot{\left\langle a\right\rangle } & = & (-i\Delta_{a}^{\prime}-\frac{\Lambda_{a}}{2})\left\langle a\right\rangle -i(J_{-}-i\frac{\Gamma}{2}e^{-i\phi})\left\langle b\right\rangle -i\varepsilon,\nonumber \\
\dot{\left\langle b\right\rangle } & = & (-i\Delta_{b}^{\prime}-\frac{\Lambda_{b}}{2})\left\langle b\right\rangle -i(J_{+}-i\frac{\Gamma}{2}e^{i\phi})\left\langle a\right\rangle ,\label{eq:Quantum dynamics}
\end{eqnarray}
where $\Lambda_{a(b)}=\Gamma+\kappa_{a(b)}$ represent the effective decay rates of the respective cavity modes. The energy stored in the battery can alternatively be expressed as $E_{B}=\textrm{Tr}_{a}[\rho H]=\omega_{b}\langle b^{\dagger}b\rangle$, with $\textrm{Tr}_{a}[.]$ denoting the partial trace over the charger's degree of freedom. Consequently, computing the energy of each subsystem necessitates determining the second-order moments of the relevant operators.

\begin{eqnarray}
\langle\dot{a^{\dagger}a}\rangle & = & -\Lambda_{a}\left\langle a^{\dagger}a\right\rangle -2\textrm{Re}\left\{ -i(J_{-}-i\frac{\Gamma}{2}e^{-i\phi})\left\langle a^{\dagger}b\right\rangle \right\} -2\textrm{Im}\left\{ \varepsilon\left\langle a\right\rangle \right\} ,\nonumber \\
\langle\dot{b^{\dagger}b}\rangle & = & -\Lambda_{b}\left\langle b^{\dagger}b\right\rangle +2\textrm{Re}\left\{ i(J_{-}+i\frac{\Gamma}{2}e^{-i\phi})\left\langle a^{\dagger}b\right\rangle \right\} ,\nonumber \\
\langle\dot{a^{\dagger}b}\rangle & = & \left[i(\Delta_{a}^{\prime}-\Delta_{b}^{\prime})-\frac{(\Lambda_{a}+\Lambda_{b})}{2}\right]\left\langle a^{\dagger}b\right\rangle +i\varepsilon\left\langle b\right\rangle \nonumber \\
 &  & -i(J_{+}-i\frac{\Gamma}{2}e^{i\phi})\left\langle a^{\dagger}a\right\rangle +i(J_{+}+i\frac{\Gamma}{2}e^{i\phi})\left\langle b^{\dagger}b\right\rangle,\label{eq:Second-order quantum dynamics}
\end{eqnarray}
where $\langle{a^{\dagger}b}\rangle$ denotes the energy transfer between the charge and the battery. In the following, based on the physical model above, we will implement the fully nonreciprocal QB and QB at EP.

To realize non-reciprocity, precise modulation of the dissipative rates and coherent interaction strength is essential. When the condition $J_{-}=i\frac{\Gamma}{2}e^{-i\phi}$ is satisfied, unidirectional energy flow from the charger to the battery is achieved, with complete suppression of energy backflow. In this scenario, we find that $\langle a^{\dagger}a\rangle$ is independent of $\left\langle a^{\dagger}b\right\rangle $ and $\left\langle a b^{\dagger}\right\rangle $, while $\langle b^{\dagger}b\rangle$ is correlated with $\left\langle a^{\dagger}b\right\rangle$ and $\left\langle a b^{\dagger}\right\rangle $, and moreover, $\langle a^{\dagger}a\rangle$ modulates $\langle b^{\dagger}b\rangle$ indirectly through the intermediary of $\left\langle a^{\dagger}b\right\rangle$ and $\left\langle a b^{\dagger}\right\rangle$. This unique coupling mechanism ensures efficient directional energy transmission and effective energy accumulation within the QB.

We assume that the initial states of both the battery and charger are the vacuum states. Under the resonance conditions $\Delta_{a}^{\prime}=\Delta_{b}^{\prime}=0$, Eqs. (\ref{eq:Quantum dynamics}) and Eqs. (\ref{eq:Second-order quantum dynamics}) can be readily solved, giving rise to the explicit solutions (\ref{Appendix:B}):
\begin{eqnarray}
\left\langle a^{\dagger}a\right\rangle & = & \frac{4\varepsilon^{2}}{\Lambda^{2}}(1-e^{-\frac{\Lambda_{a}}{2}t})^{2}, \\ \label{eq:Second-order solutions 1}
\left\langle b^{\dagger}b\right\rangle & = & \frac{16 \varepsilon^{2}\Gamma^{2}}{\varLambda_{a}^{2}}[\frac{A_{1}}{\varLambda_{b}}(1-e^{-\varLambda_{b}t})\nonumber 
+\frac{A_{2}}{\varLambda_{b}-\frac{\varLambda_{a}}{2}}(e^{-\frac{\varLambda_{a}}{2}t}-e^{-\varLambda_{b}t})\nonumber \\ \label{eq:Second-order solutions 2}
 &  & +\frac{A_{3}}{\varLambda_{b}-\varLambda_{a}}(e^{-\varLambda_{a}t}-e^{-\varLambda_{b}t})\nonumber +\frac{2A_{4}}{\varLambda_{b}}(e^{-\frac{\varLambda_{b}}{2}t}-e^{-\varLambda_{b}t})\nonumber \\
 &  & -2\frac{A_{1}+A_{2}+A_{3}+A_{4}}{\left(\varLambda_{b}-\varLambda_{a}\right)}(e^{-\frac{\varLambda_{a}+\varLambda_{b}}{2}t}-e^{-\varLambda_{b}t})],\\
\left\langle a^{\dagger}b\right\rangle  & = & -\frac{8\varepsilon^{2}\Gamma e^{i\phi}}{\varLambda_{a}^{2}}\text{[}A_{1}\left(1-e^{-\frac{\varLambda_{a}+\varLambda_{b}}{2}t}\right)
 +A_{2}\left(e^{-\frac{\varLambda_{a}}{2}t}-e^{-\frac{\varLambda_{a}+\varLambda_{b}}{2}t}\right)\nonumber \\
 &  & +A_{3}\left(e^{-\varLambda_{a}t}-e^{-\frac{\varLambda_{a}+\varLambda_{b}}{2}t}\right)
 +A_{4}\left(e^{-\frac{\varLambda_{b}}{2}t}-e^{-\frac{\varLambda_{a}+\varLambda_{b}}{2}t}\right)]\text{.} \label{second ab}
\end{eqnarray}

with{\small
\begin{eqnarray}
A_{1} & = & \frac{1}{\varLambda_{a}+\varLambda_{b}}+\frac{\varLambda_{a}}{\varLambda_{b}\left(\varLambda_{a}+\varLambda_{b}\right)},\nonumber \\
A_{2} & = & -\frac{2}{\varLambda_{b}}+\frac{\varLambda_{a}}{\varLambda_{b}\left(\varLambda_{a}-\varLambda_{b}\right)},\nonumber \\
A_{3} & = & -\frac{1}{\left(\varLambda_{a}-\varLambda_{b}\right)},\nonumber \\
A_{4} & = & -\frac{\varLambda_{a}}{\varLambda_{b}\left(\varLambda_{a}-\varLambda_{b}\right)}.\nonumber
\end{eqnarray}
}
In the long-time limit $t\to\infty$, the steady-state energies stored in the charger and the battery are
\begin{eqnarray}
E_{A}(\infty) & = & \frac{4\omega_{a}\varepsilon^{2}}{\Lambda_{a}^{2}},\nonumber \\
E_{B}(\infty) & = & \frac{16\omega_{b}\varepsilon^{2}\Gamma^{2}}{\Lambda_{a}^{2}\Lambda_{b}\left(\Lambda_{a}+\Lambda_{b}\right)}+\frac{16\omega_{b}\varepsilon^{2}\Gamma^{2}}{\Lambda_{a}\Lambda_{b}^{2}\left(\Lambda_{a}+\Lambda_{b}\right)}.\label{eq:Steady-state enrgy}
\end{eqnarray}
To evaluate the performance of a QB, we define a key metric: the charger-to-battery energy transfer efficiency over a specified time interval, which can be given by
\begin{equation}
\eta(t)=\frac{E_{B}}{E_{A}+E_{B}},\label{eq:Instantaneous charging power-1-1}
\end{equation}
It should be noted that while the nonreciprocal behavior and the final steady-state outcomes are independent of the initial state, the total charging time may be affected. To analyze this, we further define the instantaneous charging power of the QB as
\begin{equation}
P(t)=\frac{\partial E_B(t)}{\partial t}.\label{eq:Instantaneous charging power-1}
\end{equation}
As the system approaches steady state, the charging energy saturates, and the power flow $P(t)$ correspondingly tends to zero.

\section{Results and discussions}

\begin{figure}
    \centering
           \includegraphics[width=0.8\columnwidth]{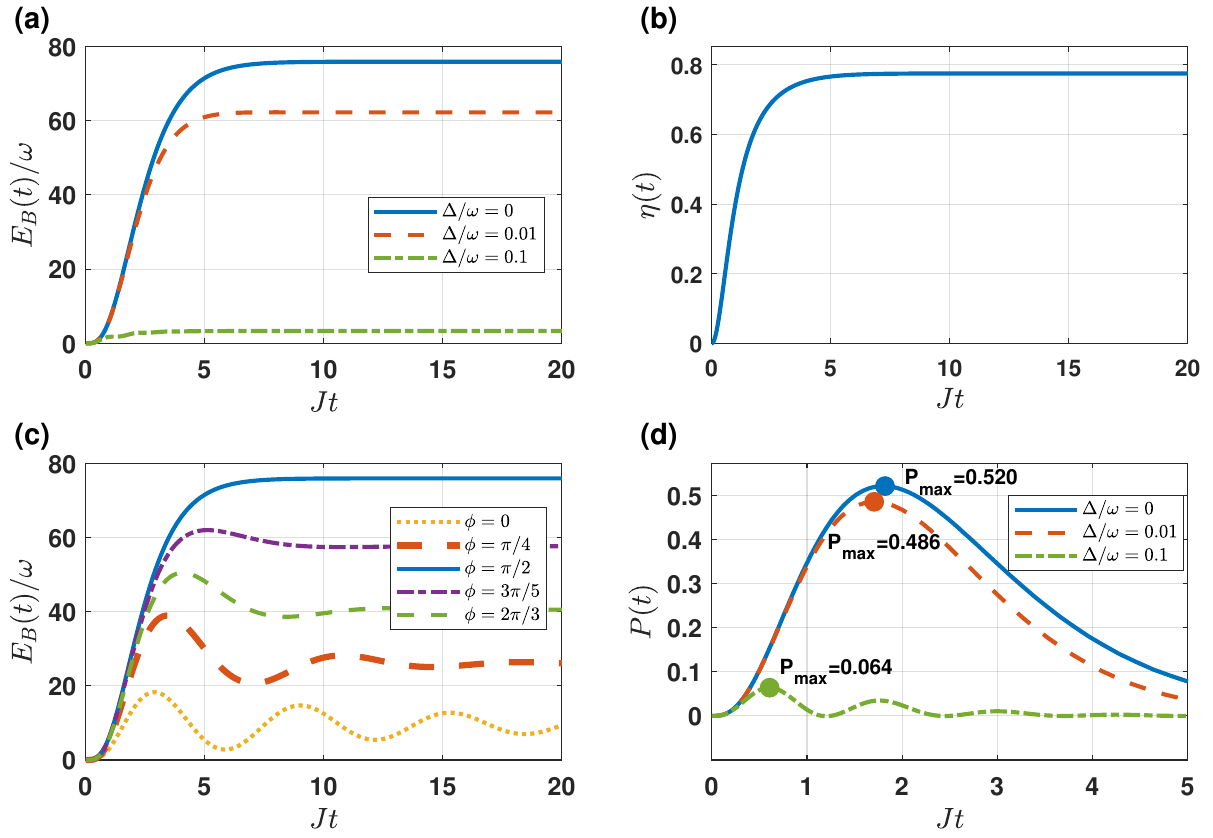}
\caption{\label{fig:2}(a) The time evolution of the battery energy $E_{B}$ versus detuning $\Delta$. (b) The time evolution of the charger-battery energy transfer efficiency $\eta(t)$. (c)$E_{B}(t)$ for different phases $\phi$. (d)The instantaneous charging power $P(t)$ for different $\Delta$. The relevant parameters are $\Delta_{a}^{\prime}=0$, $\varepsilon=0.1\omega$, $\phi=\pi/2$, $\Gamma_{a}=\Gamma_{b}=\Gamma=0.04\omega$, $\kappa_{a}=\kappa_{b}=0.003\omega$, $\gamma=10\omega$, and $J=\Gamma/2$.}
\end{figure}

We first focus on the resonance condition between the charger and the battery, and then numerically investigate the effects of the frequency detuning, phase, and the system dissipation on the energy transfer characteristics of the QB. 

\subsection{Implementation of QB} 

Firstly, we fix the parameters as $\Gamma_{a}=\Gamma_{b}=\Gamma=0.04\omega$, $\kappa_{a}=\kappa_{b}=\kappa=0.003\omega$, $\varepsilon=0.1\omega$ and $\gamma=10\omega$ to satisfy the nonreciprocal energy transfer conditions. Considering  $\Delta_a' = 0$,  we investigate the effect of frequency detuning $\Delta=\Delta_b' - \Delta_a'$ on the battery energy $E_B$ through numerical simulations. The results reveal a clear trend: $E_{B}(\Delta=0)>E_{B}(\Delta\neq0)$. As shown in Fig. \ref{fig:2}(a), the steady-state storage energy of the battery reaches $E_B=75.9$ for $\Delta=0$ (the blue solid line), while it reduces to  $E_B=62.2$ for $\Delta=0.01\omega$ (the red dotted line). Resonant coupling minimizes the energy mismatch between the two modes. This allows the charger to efficiently transfer energy to the battery without significant loss to off-resonant dissipation channels. This behavior is consistent with the nonreciprocal condition, which relies on tuned coherent coupling to suppress backflow. When $\Delta=0.1\omega$ (the green dot-dashed line), the battery energy will drop to $3.3$. The sharp performance drop results from large detuning, which breaks resonance, suppresses coherent energy exchange, and leads most of the charger energy to dissipate irreversibly instead of being stored in the battery. Consequently, high-efficiency QB operation requires the detuning to be kept as small as possible. \label{frequency detuning}

Under nonreciprocal conditions, the steady-state value ratio $E_B/E_A =4\Gamma^{2}/\Lambda^{2}\thickapprox 4$. The steady-state energy transfer efficiency $\eta(\infty)\thickapprox 0.78$ is shown in Fig. \ref{fig:2}(b). $\Gamma > \kappa$ results in the battery energy exceeding that of the charger. This is because a larger $\Gamma$ signifies a stronger coupling of the battery to the environment, enhancing its energy-exchange capacity. By reducing energy leakage via a smaller $\kappa$, the system channels a greater share of the net energy flow into the battery. This results in the battery energy exceeding that of the energy-supplying charger. 

In NH systems, phase parameters often modulate the coupling symmetry between subsystems. The phase $\phi=\pi/2$ breaks the reciprocity of energy exchange, which ensures that energy flow is directed toward the battery and suppressed in the reverse direction. However, when phase angles deviating from $\pi/2$ are adopted, it is observed that the battery energy decreases, accompanied by temporal fluctuations in energy before the steady state is reached. Moreover, the time necessary to attain the steady state is extended, as shown in Fig. \ref{fig:2}(c). When the system fails to satisfy the nonreciprocal condition, the battery energy will exhibit a backflow phenomenon. This backflow further induces energy exchange between the battery and the charger. The battery energy shows time-dependent oscillations and finally tends to a steady state.

Furthermore, we exhibit the dynamical behavior of the instantaneous charging power $P(t)$ within the time scale $Jt<5$ in Fig. \ref{fig:2}(d). When $\Delta=0$ (the blue solid line), the energy level matching eliminates the 'energy barrier' for inter-subsystem transfer, enabling $P(t)$ to reach its peak value 0.52 at $Jt\thickapprox2$. Under the same conditions, $\eta$ simultaneously reaches $75\%$, which is close to its upper limit.  The QB tends to be saturated, so the charging power decreases with the charging time $Jt>2$. In contrast, for $\Delta=0.01\omega$ (the red dotted line), the peak value decreases to 0.486. The reduction results from energy dissipation induced by detuning, which prevents a portion of the charger's energy from reaching the QB. At larger detuning $\Delta = 0.1\omega$ (green dot-dashed line), the peak value of $P(t)$ is further reduced to 0.06. In this latter regime, $P(t)$ exhibits pronounced oscillatory behavior, resulting from intermittent energy transfer during the charging process \cite{10.1088/13672630/17/7/075015_2015}.

\subsection{The optimal damping rate}

\begin{figure}
    \centering
          \includegraphics[width=0.8\columnwidth]{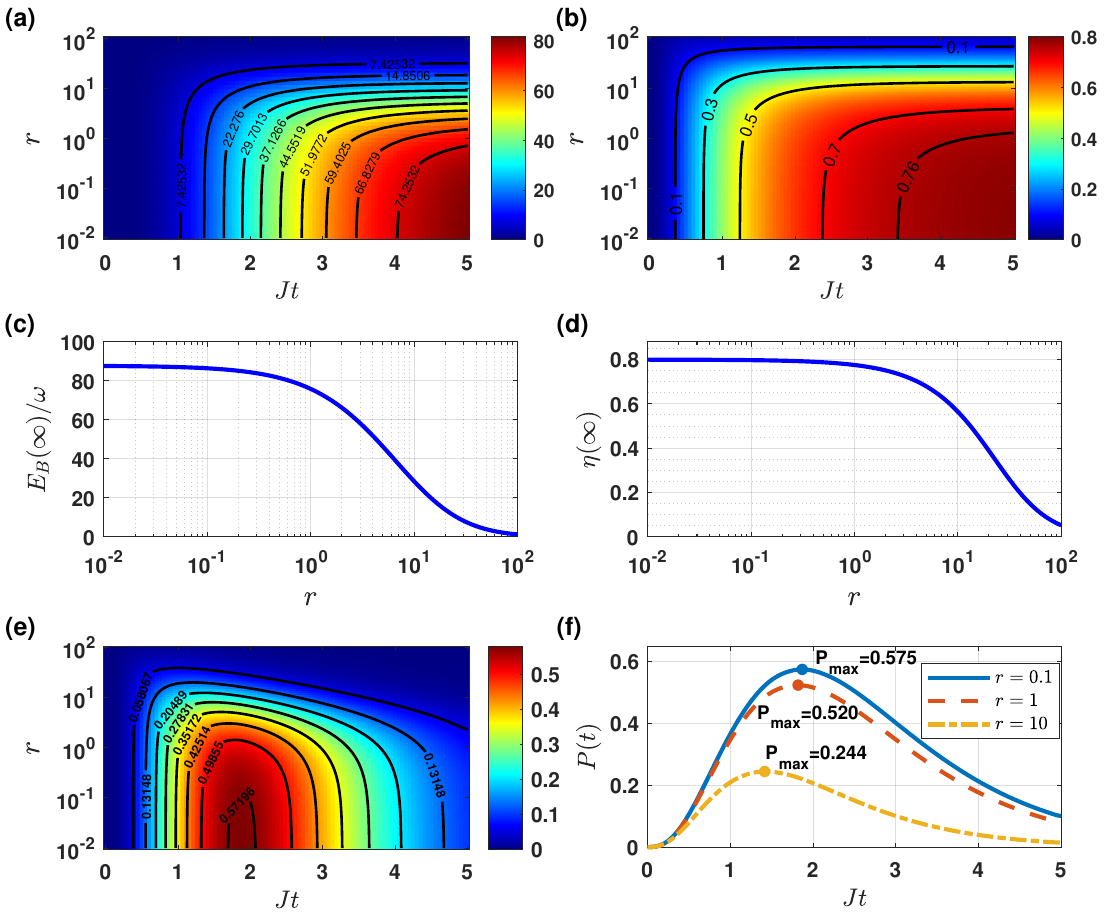}

\caption{\label{fig:3}(a)The battery energy $E_{B}(t)$ and (b) $\eta(\infty)$ versus $r=\kappa_{b}/\kappa_{a}$ and  the scaled time $Jt$.  (c) and (d) The  steady-state battery energy $E_{B}(\infty)$ and $\eta(\infty)$ for different damping ratios $r$.  (e) The instantaneous charging power $P(t)$ versus $r$ and $Jt$.  (f)The instantaneous charging power $P(t)$ for specific damping ratios $r=0.1$ (blue solid line), $r=1$ (red dashed line), $r=10$ (yellow dot-dashed line).  Other parameters are consistent with Fig. 2.}
\end{figure}

We further investigate the damping ratio that yields the maximum energy, energy transmission efficiency, and charging power. Fixing $\kappa_{a} =0.003\omega$ and defining the damping ratio as $r=\kappa_{b}/\kappa_{a}$, we plot the time-dependent battery energy $E_{B}(t)$ and energy transmission efficiency $\eta(t)$ as functions of $r$ and $Jt$ in Figs. \ref{fig:3}(a) and \ref{fig:3}(b). Reducing $r$ enhances both $E_{B}(t)$ and $\eta(t)$: for a fixed charging time $Jt=4$, decreasing $r$ from $10$ to $0.01$ boosts the battery energy from $14.8$ to $87$. Physically, this behavior arises because the final battery energy increases when the energy loss from the battery to the environment is smaller than that from the charger to the environment. 

Additionally, Fig. \ref{fig:3}(c) and Fig. \ref{fig:3}(d) depict the steady-state quantities $E_{B}(\infty)$ and $\eta(\infty)$ versus $r$, respectively. We find that when $r>100$, $\eta(\infty)$ becomes extremely low, precluding effective charging of the QB. In this regime, $\kappa_{b}$ is vastly larger than $\kappa_{a}$, meaning the battery acts as a high-loss channel that cannot retain any energy transferred from the charger, even as the charger itself maintains a low-dissipation state. As $r$ decreases from $100$ to $1$, both $\eta(\infty)$ and $E_{B}(\infty)$ improve drastically, as the dissipation rate of the battery is brought into closer alignment with that of the charger, optimizing the nonreciprocal energy transfer condition-this alignment minimizes the backflow of energy from the battery to the charger, which is a key bottleneck in conventional reciprocal QB architectures. For $r < 1$, both $\eta(\infty)$ and $E_B(\infty)$ exhibit a marked deceleration in their increase, ultimately converging to a steady state. Notably, in the vicinity of $r=1$, both the energy and the transmission efficiency of the nonreciprocal QB display remarkable robustness against fluctuations in $r$. This is a critical feature for experimental implementation, as it arises from the flatness of the performance landscape around the balanced dissipation point ($r=1$), where small deviations in $\kappa_{a}$ or $\kappa_{b}$ do not disrupt the delicate nonreciprocal coupling that enables efficient energy transfer.

Finally, the influence of the damping ratio on the maximum short-time instantaneous charging power is investigated. Fig. \ref{fig:3}(e) illustrates the variation of $P(t)$ with respect to the damping ratio $r$ and the scaled time $Jt$. When the scaled time $Jt$ ranges from $0.5$ to $3.5$ and the damping ratio $r<3$, the color block representing $P(t)$ transitions from blue (inefficient regime, $P<0.1$) to red (efficient regime, $P\thickapprox 0.58$). Physically, this transition stems from balancing two competing effects via dissipation engineering. Decreasing $r$ enhances the battery energy retention, while higher $\kappa_{a}$ accelerates unidirectional energy transfer, which is consistent with the nonreciprocal condition that suppresses backflow. Under such optimized parameters, $P(t)$ nears its peak, enabling a shift from low-power to high-power regimes. This validates the dissipation role of the engineering in boosting power and provides critical parameter constraints for self-optimizing quantum power chips \cite{10.1103/PhysRevX.8.031007_2018}. It is worth noting that the charging power $P$ increases in tandem with the damping ratio. The maximum instantaneous power remains relatively stable when $r\leq1$, as shown in Fig. \ref{fig:3}(f). This stability originates from the flat performance landscape around the balanced dissipation point ($r=1$), where small fluctuations in $\kappa_{a}$ or $\kappa_{b}$ do not disrupt the optimal nonreciprocal coupling. In the regime where $r\leq1$, $\kappa_{a}$ ceases to limit the system, and $\kappa_{b}$ is sufficiently low to prevent energy loss, thereby creating a sweet spot for stable high-power operation.

By optimizing the dissipation ratio of the charger and battery based on these insights, one can further enhance the charging efficiency, as the dissipation ratio directly tunes the strength of nonreciprocal energy transfer and the balance between energy injection and retention.

\section{QB at EP}

\begin{figure}
    \centering
     \includegraphics[width=0.8\columnwidth]{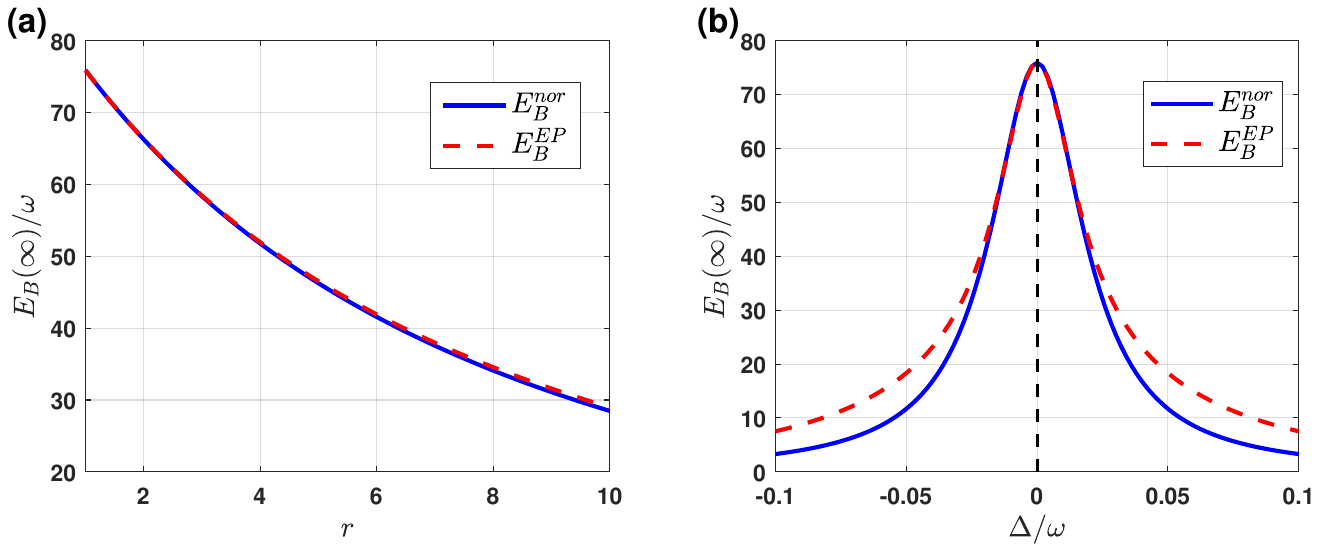}

\caption{\label{fig:4}(a) The battery energy in the case of non-reciprocity ( blue solid line) and QB at EP (red dashed line) versus damping ratio $r$ for $\Delta=0$, and (b) versus $\Delta$ with $r=1$. Other parameters are consistent with Fig. 2.}
\end{figure}

It has been shown that the NH system working at an EP can enhance energy transfer efficiency of the QB by tuning the coherent-dissipative coupling phase and dissipation ratio \cite{arXiv:2504.15091_2025,10.1103/PhysRevLett.133.180801_2024, 
10.1038/s41467024534348_2024}. We seek to achieve the QB at EP in our NH model. According to the effective Hamiltonian Eq.  (\ref{eq:Quantum dynamics}), the dynamical equations can be written in a matrix form:
\begin{eqnarray}
\frac{d}{dt}\begin{pmatrix}\langle a\rangle\\
\langle b\rangle
\end{pmatrix} & = & M\begin{pmatrix}\langle a\rangle\\
\langle b\rangle
\end{pmatrix}+\begin{pmatrix}-i\varepsilon\\
0
\end{pmatrix},
\end{eqnarray}
where $M=\begin{pmatrix}A & B\\
C & D
\end{pmatrix}$ is the dynamical matrix, $A=-i\Delta_{a}^{\prime}-\Lambda_{a}/2$, $B=-iJ_{-}-(\Gamma/2)e^{-i\phi}$, $C=-iJ_{+}-(\Gamma/2) e^{i\phi}$, $D=-i\Delta_{b}^{\prime}-\Lambda_{b}/2$. Its eigenvalues determine the system's transient behavior (e.g., decay rates). Fixing the driving strength $\varepsilon=0$, the eigenvalues $\lambda$ satisfy the characteristic equation $\det(M-\lambda I)=0$. The eigenvalues $\lambda$ are given by
\begin{eqnarray}
\lambda_{\pm} & = & i\frac{\Delta_{a}^{'}+\Delta_{b}^{'}}{2}+(\frac{\Lambda_{a}+\Lambda_{b}}{4})\nonumber \\
& & \pm \frac{1}{2}\sqrt{\left[-i\left(\Delta_{a}^{\prime}-\Delta_{b}^{\prime}\right)-\frac{\varLambda_{a}-\varLambda_{b}}{2}\right]^{2}+4(-iJ_{-}-\frac{\Gamma}{2}e^{-i\phi})(-J_{+}-\frac{\Gamma}{2}e^{i\phi})}.
\end{eqnarray}

An exceptional point requires that the eigenvalues and eigenvectors of $M$ are degenerate, which is equivalent to the discriminant being zero, i.e.
{\small
\begin{eqnarray}
\left[-i\left(\Delta_{a}^{\prime}-\Delta_{b}^{\prime}\right)-\frac{\varLambda_{a}-\varLambda_{b}}{2}\right]^{2}+4(-iJ_{-}-\frac{\Gamma}{2}e^{-i\phi})(-J_{+}-\frac{\Gamma}{2}e^{i\phi})=0.\label{EP}
\end{eqnarray}
}
At an exceptional point, the system exhibits significantly enhanced sensitivity to external driving. Such enhancement, when applied to a QB operating near an EP, can amplify the steady-state responses $\langle a\rangle_{ss}$ and $\langle b\rangle_{ss}$, thereby increasing the stored energy. We find that a QB operating at EP corresponds to a partially nonreciprocal regime, characterized by unequal off-diagonal elements in the effective coupling matrix $M$. 

We consider two distinct parameter regimes: (i) $\Delta = 0$, $\phi = \pi/2$ with varying damping ratio $r$, and (ii) $r = 1$, $\phi = \pi/2$ with varying  $\Delta$. The corresponding results are presented in Fig. \ref{fig:4}(a) and (b), respectively, where $E^{\text{nor}}$ and $E^{EP}$  denote battery energy in the full nonreciprocal regime and the EP condition, respectively. Fig. \ref{fig:4}(a) shows that a smaller $r$ leads to a higher steady-state energy, $E^{\text{nor}}$ and $E^{EP}$  exhibit closely matched trends without significant differences. Fig. \ref{fig:4}(b) shows that  $E^{nor}$ and $E^{EP}$ coincide only at $\Delta = 0$, corresponding to the best working point for both regimes. Although $E^{\text{nor}}$ and $E^{EP}$  maintain comparable QB energy under small detunings, the stored energy decreases rapidly with increasing $|\Delta |$, which weakens the coherent cavity coupling by modifying the off-diagonal Hamiltonian terms. While the completely nonreciprocal condition $J=|\Gamma/2|$ is independent of  $\Delta$, the QB working at EP requires adjusting the ratio $J/\Gamma$ when $\Delta$ varies to maintain the exceptional point. This adjustment introduces a synergistic effect that partially counteracts the detrimental influence of detuning. As a result, under any finite detuning, the QB working at EP consistently exhibits higher energy retention than the nonreciprocal QB.

\section{Experimental Feasibility Discussion}

In this section, we explain the experimental realization of our scheme. The effective physical model requires $\gamma_{m}\gg\{\kappa_{a},\kappa_{b}\}$, $\left|\gamma_{m}/2\right|\gg\{g_{a},g_{b}\}$ and $J=\left|\Gamma/2\right|$. The current proposal can be implemented using organic microcavities\cite{10.1126/sciadv.abk3160} or superconducting technology \cite{10.1038/s41467024546628_2024, 10.1088/2058-9565/ac8444}. The NH three-mode system has been realized in a circuit quantum electrodynamics system, including a Josephson-junction-based qubit (battery), a bus resonator (charge), and a readout resonator (shared mode). The decay rates of charge and battery are respectively $\kappa_{a}/2\pi=0.08$ MHz and $\kappa_{b}/2\pi=0.06$ MHz, and that of the shared mode $\gamma_{m}/2\pi=5$ MHz \cite{10.1038/s41467024546628_2024}. 
We use a precision-tunable coupler, with the corresponding coupling strength ranging from $0$ MHz to $123$ MHz  \cite{10.1103/PhysRevApplied.19.064043_2023}. The coupling strength can be engineered as $g_{a}/2\pi=g_{b}/2\pi\approx 0.33$ MHz and $J/2\pi=0.01$ MHz. Consequently, the dissipative coupling strength is obtained as $\Gamma/2\pi=g_{a}^{2}/\gamma_{m}\approx 0.02$ MHz.

The proposed scheme can also be implemented on a range of experimental platforms. Optical systems using quantum switches have enabled the realization of similar directional processes \cite{PhysRevLett.131.240401, 10.1103/PhysRevLett.131.260401, 10.1364/OL.446367}. Alternatively, the model could be implemented in solid-state systems, such as the electron spin of a nitrogen-vacancy (NV) center in diamond,  which is supported by an ancillary nuclear spin \cite{PhysRevLett.133.180401}. Energy injection and extraction in molecular structures composed of a central battery spin surrounded by charger spins have also been demonstrated using nuclear magnetic resonance techniques \cite{10.1103/PhysRevA.106.042601}. Moreover, recent progress in room-temperature QBs, such as those based on multi-layered organic microcavities, marks a key step toward realizing the full operational cycle of a QB \cite{arXiv:2501.16541}. These diverse experimental approaches support the practical feasibility of our scheme.

\section{Conclusion}

In summary, we have proposed an NH scheme to achieve a nonreciprocal QB. By employing a strongly dissipative cavity as an auxiliary reservoir, we enable an effective indirect coupling between the charger and the battery. In the fully nonreciprocal case, i.e., under the destructive interference condition, we find that the energy of the battery exceeds that of the charger, provided that the couplings of both the charger and the battery to the common reservoir are stronger than the local damping of each mode. Moreover, the optimal working point occurs at resonance with a small $\kappa_{b} / \kappa_{a}$ ratio. Energy storage efficiency improves when the system operates at the EP, forming a partially nonreciprocal case. Compared to the fully nonreciprocal case, the partially nonreciprocal QB is more robust to parameter fluctuations. Our work may provide opportunities for directional energy transfer, controlled dissipation, and entropy management in open quantum systems \cite{10.1103/RevModPhys.96.031001_2024}.

% This section is a list of funder names and grant numbers

\ack{Sample text inserted for demonstration.}

\funding{This work is supported by the National Natural Science Foundation of China under Grants No. 12174058.}
% This section is a list of funder names and grant numbers

\roles{Sample text inserted for demonstration.}
% List author names and the contributions made to the article, using terms from the NISO Contributor Roles Taxonomy (CRediT) https://credit.niso.org

\data{Sample text inserted for demonstration.}
% For more information on IOP Publishing's research data policy see: https://publishingsupport.iopscience.iop.org/questions/research-data/

\suppdata{Sample text inserted for demonstration.}

\appendix

\renewcommand{\thesection}{Appendix: \Alph{section}} 

\section{Derivation Of The Effective Hamiltonian}\label{Appendix:A}

Based on the coupling and dissipation properties of the established three-bosonic-mode model, the Hamiltonian of the system is written as follows:
\begin{eqnarray}
H & = & \Delta_{a}a^{\dagger}a+\Delta_{b}b^{\dagger}b+(\Delta_{c}-i\frac{\gamma_{m}}{2})c^{\dagger}c\nonumber \\
 &  & +g_{a}e^{i\phi}ac^{\dagger}+g_{a}e^{-i\phi}a^{\dagger}c+g_{b}(bc^{\dagger}+b^{\dagger}c)\nonumber \\
 &  & +J(a^{\dagger}b+ab^{\dagger})+\varepsilon(a+a^{\dagger}).\label{eq:The system Hamiltonian}
\end{eqnarray}
We first define the projection operator $P=\sum_{n}(|a_{n}\rangle\langle a_{n}|+|b_{n}\rangle\langle b_{n}|)$ and $Q=\sum_{n}|c_{n}\rangle\langle c_{n}|$, where $P+Q=I$, $P^{2}+Q^{2}=I$. The projected Hamiltonian within the subspace $\{|a_{n}\rangle,|b_{n}\rangle\}$ is then
\begin{eqnarray}
H_{eff} & = & PHP+\frac{PHQQHP}{E-QHQ},
\end{eqnarray}
where $H$ is from Eq. (\ref{eq:The system Hamiltonian}). For $E\ll QHQ\sim|\Delta_{c}-i\frac{\gamma_{m}}{2}|$,  it follows 
\begin{eqnarray}
H_{eff} & \approx & \Delta_{a}a^{\dagger}a+\Delta_{b}b^{\dagger}b+\varepsilon\left(a+a^{\dagger}\right)+Ja^{\dagger}b+Jab^{\dagger}\nonumber \\
 &  & +\frac{g_{a}^{2}a^{\dagger}a+g_{b}^{2}b^{\dagger}b+g_{a}g_{b}e^{-i\phi}a^{\dagger}b+g_{a}g_{b}e^{i\phi}ab^{\dagger}}{-\left(\Delta_{c}-i\gamma\right)}\nonumber \\
 & = & \left(\Delta_{a}^{\prime}-i\Gamma_{a}\right)a^{\dagger}a+\left(\Delta_{b}^{\prime}-i\Gamma_{b}\right)b^{\dagger}b+ \varepsilon\left(a+a^{\dagger}\right)\nonumber \\
 &  & +\left[J-\left(G+i\Gamma\right)e^{-i\phi}\right]a^{\dagger}b\nonumber \\
 &  & +\left[J-\left(G+i\Gamma\right)e^{i\phi}\right]ab^{\dagger}, \label{A17}
\end{eqnarray}
where $\frac{\gamma_{m}}{2}=\gamma$,  $\Delta_{a}^{\prime}=\Delta_{a}-\frac{g_{a}^{2}\Delta_{c}}{\Delta_{c}^{2}+\gamma^{2}}$, $\Delta_{b}^{\prime}=\Delta_{b}-\frac{g_{b}^{2}\Delta_{c}}{\Delta_{c}^{2}+\gamma^{2}}$,
$\Gamma_{a}=\frac{g_{a}^{2}\gamma}{\Delta_{c}^{2}+\gamma^{2}}$, $\Gamma_{b}=\frac{g_{b}^{2}\gamma}{\Delta_{c}^{2}+\gamma^{2}}$,
$G=\frac{g_{a}g_{b}\Delta_{c}}{\Delta_{c}^{2}+\gamma^{2}}$, $\Gamma=\frac{\gamma g_{a}g_{b}}{\Delta_{c}^{2}+\gamma^{2}}$,
and $J_{\pm}=J-Ge^{\pm i\phi}$. We obtain the effective Hamiltonian as Eq. (\ref{eq:Effective Hamiltonian}) in the main text. Through this method above, the mode $c$ is adiabatically eliminated. The entire system can be regarded as the interaction between modes, including coherent and dissipative coupling interactions. 

\section {Derivation Of Subsystem Energies}\label{Appendix:B}

Based on the derived effective Hamiltonian Eq. (\ref{A17}), we proceed with further analysis. We assume that the reservoir is Markovian and start from the Lindblad master equation:  $\dot{\rho}=-i[H_\textrm{eff},\rho]+\sum_{j=a,b}\kappa_{j}\mathcal{L}[j]\rho$, where $\mathcal{L}[o]\rho=o\rho o^{\dagger}-\frac{1}{2}\left\{ o^{\dagger}o,\rho\right\} $.  Then, we get the evolution of the operators governed by the following equations:
\begin{eqnarray}
\dot{\left\langle a\right\rangle } & = & \left(-i\Delta_{a}^{\prime}-\frac{\Gamma_{a}}{2}-\frac{\kappa_{a}}{2}\right)\left\langle a\right\rangle +\left(-iJ_{-}-\frac{\Gamma}{2}e^{-i\phi}\right)\left\langle b\right\rangle -i\varepsilon,\nonumber \\
\dot{\left\langle b\right\rangle } & = & \left(-i\Delta_{b}^{\prime}-\frac{\Gamma_{b}}{2}-\frac{\kappa_{b}}{2}\right)\left\langle b\right\rangle +\left(-iJ_{+}-\frac{\Gamma}{2}e^{i\phi}\right)\left\langle a\right\rangle ,\\
\langle\dot{a^{\dagger}a}\rangle & = & -\left(\Gamma_{a}+\kappa_{a}\right)\left\langle a^{\dagger}a\right\rangle -iJ_{-}\left\langle a^{\dagger}b\right\rangle +iJ_{+}\left\langle ab^{\dagger}\right\rangle \nonumber \\
 &  & -\frac{\Gamma}{2}e^{i\phi}\left\langle ab^{\dagger}\right\rangle -\frac{\Gamma}{2}e^{-i\phi}\left\langle a^{\dagger}b\right\rangle -2\varepsilon Im\left\{ \left\langle a\right\rangle \right\} ,\label{eq: a mode polulation}\\
\langle\dot{b^{\dagger}b}\rangle & = & -\left(\Gamma_{b}+\kappa_{b}\right)\left\langle b^{\dagger}b\right\rangle +iJ_{-}\left\langle a^{\dagger}b\right\rangle -iJ_{+}\left\langle ab^{\dagger}\right\rangle \nonumber \\
 &  & -\frac{\Gamma}{2}e^{i\phi}\left\langle ab^{\dagger}\right\rangle -\frac{\Gamma}{2}e^{-i\phi}\left\langle a^{\dagger}b\right\rangle ,\nonumber \\
\langle\dot{a^{\dagger}b}\rangle & = & \left[i\left(\Delta_{a}^{\prime}-\Delta_{b}^{\prime}\right)-\frac{1}{2}\left(\kappa_{a}+\kappa_{b}+\Gamma_{a}+\Gamma_{b}\right)\right]\left\langle a^{\dagger}b\right\rangle +i\varepsilon\left\langle b\right\rangle \nonumber \\
 &  & +\left(-iJ_{+}-\frac{\Gamma}{2}e^{i\phi}\right)\left\langle a^{\dagger}a\right\rangle +\left(iJ_{+}-\frac{\Gamma}{2}e^{i\phi}\right)\left\langle b^{\dagger}b\right\rangle.\nonumber 
\end{eqnarray}
Assumed $g_{a}=g_{b}=g$, $\Lambda_{a}=\Gamma+\kappa_{a}$, and $\Lambda_{b}=\Gamma+\kappa_{b}$,  we get $G=0$, $J_{-}=J_{+}^{*}$  (since $G=0$ and $\Gamma$ is real). We obtain the quantum dynamical equation as Eq. (\ref{eq:Quantum dynamics}) in the main text. Under nonreciprocal conditions $J_{-}=i(\Gamma/2)e^{-i\phi}$, $\left\langle a\right\rangle $ becomes independent of the mode $b$.

Then
\begin{eqnarray}
\left\langle a\right\rangle & = & e^{\left(-i\Delta_{a}^{\prime}-\frac{\varLambda_{a}}{2}\right)t}\left[\left\langle a\right\rangle (0)+\int_{0}^{t}e^{\left(i\Delta_{a}^{\prime}+\frac{\varLambda_{a}}{2}\right)t}\left(-i\varepsilon\right)dt\right]\nonumber \\
 & = & \frac{\varepsilon\left(\Delta_{a}^{\prime}-i\frac{\varLambda_{a}}{2}\right)}{\left(\Delta_{a}^{\prime}\right)^{2}+\left(\frac{\varLambda_{a}}{2}\right)^{2}}\left(1-e^{-\frac{\varLambda_{a}}{2}t}e^{-i\Delta_{a}^{\prime}t}\right),
\end{eqnarray}
with the initial condition $\left\langle a\right\rangle(0)=0$. According to the dynamical equation for $\langle b\rangle$, and taking  $\Delta_{a}^{\prime}=\Delta_{b}^{\prime}=0$, we get 
\begin{eqnarray}
\left\langle b\right\rangle  & = & \frac{4i\varepsilon\Gamma e^{i\phi}}{\varLambda_{a}}\left(\frac{1-e^{-\frac{\varLambda_{b}}{2}t}}{\varLambda_{b}}+\frac{e^{-\frac{\varLambda_{a}}{2}t}-e^{-\frac{\varLambda_{b}}{2}t}}{\varLambda_{a}-\varLambda_{b}}\right),
\end{eqnarray}
substituting $\left\langle a\right\rangle$ into Eq. (\ref{eq: a mode polulation}), and then combining and simplifying, $\left\langle a^{\dagger}a\right\rangle$  is given
\begin{eqnarray*}
\left\langle a^{\dagger}a\right\rangle & = & \frac{\varepsilon^{2}}{\left(\Delta_{a}^{\prime}\right)^{2}+\left(\frac{\varLambda_{a}}{2}\right)^{2}}\left[1+e^{-\varLambda_{a}t}-2e^{-\frac{\varLambda_{a}}{2}t}\cos\left(\Delta_{a}^{\prime}t\right)\right].
\end{eqnarray*}
{\footnotesize\par}

Similarly, taking the same initial conditions and assuming $\Delta_{a}^{\prime}=\Delta_{b}^{\prime}=0$, we get
\begin{eqnarray}
\left\langle a^{\dagger}b\right\rangle  & = & -\frac{8\varepsilon^{2}\Gamma e^{i\phi}}{\varLambda_{a}^{2}}\text{[}A_{1}\left(1-e^{-\frac{\varLambda_{a}+\varLambda_{b}}{2}t}\right)
 +A_{2}\left(e^{-\frac{\varLambda_{a}}{2}t}-e^{-\frac{\varLambda_{a}+\varLambda_{b}}{2}t}\right)\nonumber \\
 &  & +A_{3}\left(e^{-\varLambda_{a}t}-e^{-\frac{\varLambda_{a}+\varLambda_{b}}{2}t}\right)
 +A_{4}\left(e^{-\frac{\varLambda_{b}}{2}t}-e^{-\frac{\varLambda_{a}+\varLambda_{b}}{2}t}\right)]\text{.}
\end{eqnarray}
Therefore, the time evolution of $\langle b^{\dagger}b\rangle$
takes the form:
\begin{eqnarray}
\left\langle b^{\dagger}b\right\rangle & = &
\frac{16\varepsilon^{2}\Gamma^{2}}{\varLambda_{a}^{2}}[\frac{A_{1}}{\varLambda_{b}}(1-e^{-\varLambda_{b}t})\nonumber 
+\frac{A_{2}}{\varLambda_{b}-\frac{\varLambda_{a}}{2}}(e^{-\frac{\varLambda_{a}}{2}t}-e^{-\varLambda_{b}t})\nonumber \\
 &  & +\frac{A_{3}}{\varLambda_{b}-\varLambda_{a}}(e^{-\varLambda_{a}t}-e^{-\varLambda_{b}t})\nonumber +\frac{2A_{4}}{\varLambda_{b}}(e^{-\frac{\varLambda_{b}}{2}t}-e^{-\varLambda_{b}t})\nonumber \\
 &  & -2\frac{A_{1}+A_{2}+A_{3}+A_{4}}{\left(\varLambda_{b}-\varLambda_{a}\right)}(e^{-\frac{\varLambda_{a}+\varLambda_{b}}{2}t}-e^{-\varLambda_{b}t})],
\end{eqnarray}
with
\begin{eqnarray}
A_{1} & = & \frac{1}{\varLambda_{a}+\varLambda_{b}}+\frac{\varLambda_{a}}{\varLambda_{b}\left(\varLambda_{a}+\varLambda_{b}\right)},\nonumber \\
A_{2} & = & -\frac{2}{\varLambda_{b}}+\frac{\varLambda_{a}}{\varLambda_{b}\left(\varLambda_{a}-\varLambda_{b}\right)},\nonumber \\
A_{3} & = & -\frac{1}{\left(\varLambda_{a}-\varLambda_{b}\right)},\nonumber \\
A_{4} & = & -\frac{\varLambda_{a}}{\varLambda_{b}\left(\varLambda_{a}-\varLambda_{b}\right)},\nonumber
\end{eqnarray}
which corresponds to the subsystem energy given by Eq. (\ref{eq:Second-order solutions 2}) in the main text.

\end{document}